\documentclass[conference]{IEEEtran}
\IEEEoverridecommandlockouts

\usepackage[nocompress]{cite}
\usepackage{amsmath,amssymb,amsfonts}
\usepackage{algorithmic}
\usepackage{graphicx}
\usepackage{textcomp}
\usepackage{xcolor}
\usepackage{url}

\usepackage{booktabs}

\def\BibTeX{{\rm B\kern-.05em{\sc i\kern-.025em b}\kern-.08em
    T\kern-.1667em\lower.7ex\hbox{E}\kern-.125emX}}
\begin{document}

\title{SONIQUE: Video Background Music Generation Using Unpaired Audio-Visual Data}


\author{
\IEEEauthorblockN{
Liqian Zhang,
Magdalena Fuentes
}\\
\IEEEauthorblockA{
 MARL-IDM, New York University, New York, USA\\
    }
}

\maketitle

\begin{abstract}

We present SONIQUE, a model for generating background music tailored to video content. Unlike traditional video-to-music generation approaches, which rely heavily on paired audio-visual datasets, SONIQUE leverages unpaired data, combining royalty-free music and independent video sources. By utilizing large language models (LLMs) for video understanding and converting visual descriptions into musical tags, alongside a U-Net-based conditional diffusion model, SONIQUE enables  customizable music generation. Users can control specific aspects of the music, such as instruments, genres, tempo, and melodies, ensuring the generated output fits their creative vision. SONIQUE is open-source, with a demo available online.
\end{abstract}


\section{Introduction}
The recent availability of reasonable multimedia equipment (e.g. in smartphones) has transformed the way of storytelling. It is much easier nowadays to document or create narratives through video, and in quality. Users are sharing increasing amounts of video content in social media, and video editing tools have become the highly popular. A key aspect of the editing process is to blend in good background music. Good background music not only enriches the visual experience but also emphasizes the narrative transitions. However, it is often a challenge for content creators to find the right audio that matches their videos, because it is time consuming, it typically requires efforts in editing the music to fit the intension of the video, and because of the limitations of royalty-free music. In practice, amateur video editors often get help from websites like Pixabay \cite{pixabay_music} 
 or YouTube Studio 
to find royalty-free music. While these platforms offer various options, the royalty-free music available may not always fit the specific mood or narrative of the video. On the other hand, current audio generation models like Suno \cite{sunoai2023} can be expensive because those models are trained on private music datasets and they often require users to pay for commercial usage, which is adequate for commercial purposes but not necessarily for user-made video stories and content. 

On the other hand, music generation has recently experienced breakthroughs with the advent of diffusion models and transformers \cite{evans2024fast}. While numerous models have been developed to generate music from text prompts \cite{evans2024fast,riffusion2024,schneider2023mousai}, far fewer models focus on vision-based music generation. Moreover, those that do rely heavily on paired audio-visual data \cite{su2024v2meow,KANG2024123640,mckee2023languageguided,wang2024tiva}, face challenges due to the scarcity of such datasets, copyright constrains, and the subjective nature of music selection. 

\begin{figure*}[htb] 
\centerline{\includegraphics[width=14cm]{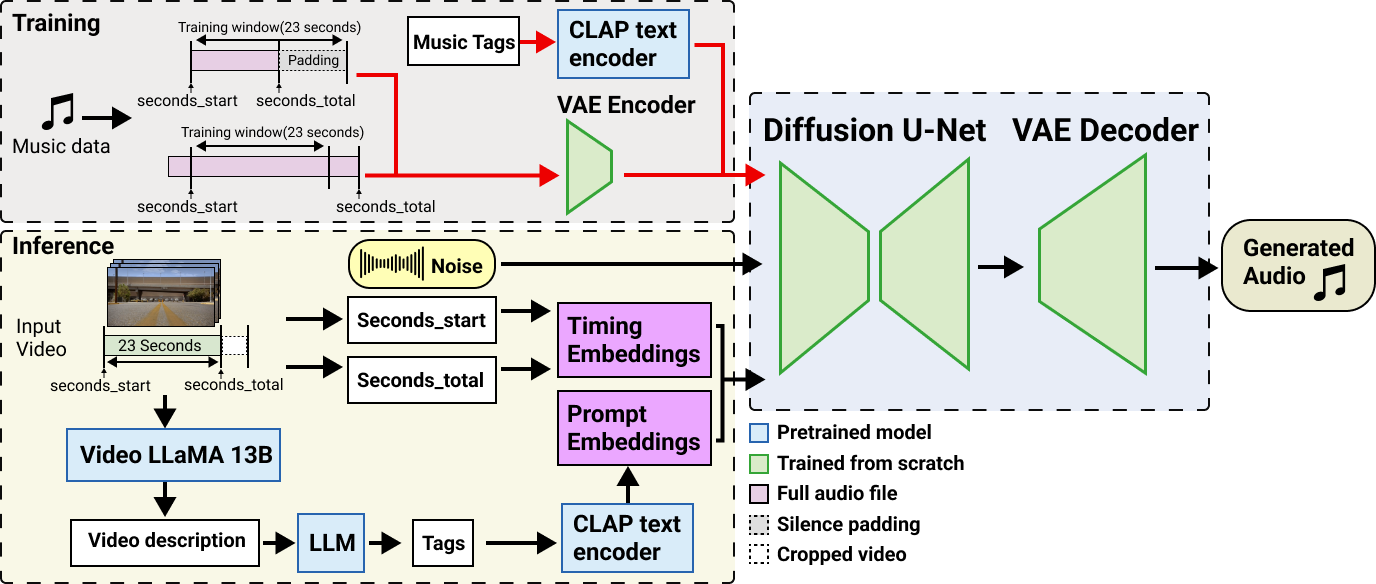}}
\caption{Our proposed SONIQUE architecture. Semantic information is extracted from the input video using Video-LLaMA \cite{zhang2023videollama}. This information is then passed through an LLM to generate simpler, descriptive tags. Then CLAP \cite{elizalde2022clap} processes these tags and any additional user-provided tags for customization to produce prompt features. Finally, a diffusion U-Net uses these features to generate background music \cite{evans2024fast}.}
\label{fig:res}
\end{figure*}

In this work, we present SONIQUE, a flexible model for video music background generation that was trained on royalty-free music and unpaired audio-visual data (i.e. videos and music that come from different sources and are not aligned). To do so we harness a combination of Large Language Models (LLMs) for video understanding and video description to tags conversion, and a recently proposed U-Net based conditional diffusion model \cite{evans2024fast}, as shown in Figure \ref{fig:res}. SONIQUE also offers users the flexibility to control the generated music by prompting the model in regards to instruments, genres, tempo rates, and melodies. SONIQUE receives a video without music and an optional prompt from the user as inputs, and creates music based on the prompt and the visuals. Our code is open source and our demo can be found online 
\footnote{\url{https://github.com/zxxwxyyy/sonique}.} 


\section{Related Work}

\noindent \textbf{Conditional Music Generation.} Among the early works on text-promped music generation, Riffusion \cite{riffusion2024} stood out as one of the first that generates spectrograms based on input text. It fine-tuned Stable Diffusion \cite{rombach2022diffusion} with images of spectrograms paired with text and produced audio from them. Similarly, Mousai \cite{schneider2023mousai} generates audio based on input text. It introduced an encoder-decoder structure that encodes compressed audio (x64) and operates with a U-Net. However, these works fell short in that 1) they lacked length control capability (i.e. the output of these models is often constrained to short and fixed lengths of audio, limiting its usability), and 2) the output audio quality falls short when compared to professional standards.

Recent work has made notable progress in addressing these challenges. Stable Audio \cite{evans2024fast} advanced the ideas from Mousai by adapting the text encoder to compress the text metadata, audio duration, and start time conditioning into the latent space. This allows control of the generation process with a predefined start time and total duration, generating longer and continuous musical pieces that vary in length based on the user's preference. Similar to Mousai, Stable Audio employs a VAE \cite{kumar2023highfidelity} autoencoder to encode input audio into latent embeddings, and the embeddings further processed through a diffusion U-Net. This approach not only improved the computational speed but also enhanced the quality and the control over the output audio. We based our model on Stable Audio \cite{evans2024fast} because it  generates high-fidelity (44.1 kHz), variable-length audio output that aligns well with the input textual description.

\noindent \textbf{Video to Audio Generation. }V2Meow \cite{su2024v2meow} encodes video into a latent representation that captures movement, color, and scene composition to generate audio. However, the reliance on paired video-audio data limits its flexibility, particularly when dealing with diverse or unpaired datasets. Additionally, V2Meow’s complex integration of motion, semantics, and discrete visual tokens from models like I3D \cite{carreira2018quo}, CLIP \cite{radford2021learning}, and ViT-VQGAN \cite{yu2022vectorquantized} can lead to challenges in producing coherent and contextually relevant audio. Similarly, Video2Music \cite{KANG2024123640} extracts semantic, motion, and emotion cues from video to generate matching music. While it uses tools such as CLIP \cite{radford2021learning} and PySceneDetect \cite{PySceneDetect} for semantic and scene analysis, its limited emotional range (e.g., 'exciting,' 'fearful,' 'tense') constrains the expressive potential of the generated music. 

McKee et al. \cite{mckee2023languageguided} propose a method for recommending music based on video and music descriptions, with user input guided by prompts. However, this approach relies on the user to have a collection of available music, and also requires the user to edit the music to fit the video (e.g. duration). Aditionally, 
its dependence on the YouTube8M dataset \cite{abuelhaija2016youtube8m} limits the quality of the generated music. Finally, Xihua et al. \cite{wang2024tiva} introduce the TiVA framework, which generates low-resolution Mel-spectrograms from video features. However, the low resolution of these spectrograms can result in a loss of temporal and sonic detail, limiting the expressiveness of the generated music. Like the other methods mentioned, TiVA also depends on paired video-audio datasets for training. 

In summary, while previous approaches rely heavily on paired video-audio datasets, limited emotional expressiveness, and/or less flexible generation mechanisms, SONIQUE’s use of unpaired data, customizable generation, and LLM-based video understanding provide a more versatile, scalable and user-friendly solution for video-to-music generation.



\section{SONIQUE}

SONIQUE's architecture is depicted in Figure \ref{fig:res}. Semantic information is extracted from the input video using Video-LLaMA \cite{zhang2023videollama}, and this is processed by a language model \cite{inferless2024mistral} to generate simpler, descriptive tags. CLAP \cite{elizalde2022clap} then processes these tags, along with any additional user-provided tags for customization, to produce prompt features. Finally, a diffusion U-Net uses these features to generate background music \cite{evans2024fast}. We only train the music encoder using lp-music-caps \cite{doh2023lpmusiccaps} to tag royalty-free music. The video input is used solely for development purposes and to inform the selection of appropriate LLMs. Each component is explained below.


\noindent \textbf{Video Understanding with LLMs.} SONIQUE utilizes Video-LLaMa \cite{zhang2023videollama}, a video understanding model to generate text description of the video. Video-LLaMA processes video input by first passing individual frames through a frozen, pre-trained image encoder. This encoder extracts  visual features from each frame, turning them into latent representations. To understand the sequence and timing of the video, positional embeddings are added to these frame representations. Next, the model uses a linear layer to convert the visual information into video query vectors. These video query vectors have the same format as text embeddings, which a frozen LLM (LLaMa) can understand. By conditioning the LLM on these video query vectors, Video-LLaMA can generate text or respond to tasks based on the video content. We prompt the model to describe the video in detail and use its output as described below.

\begin{figure*}[htb] 
\centerline{\includegraphics[width=18cm]{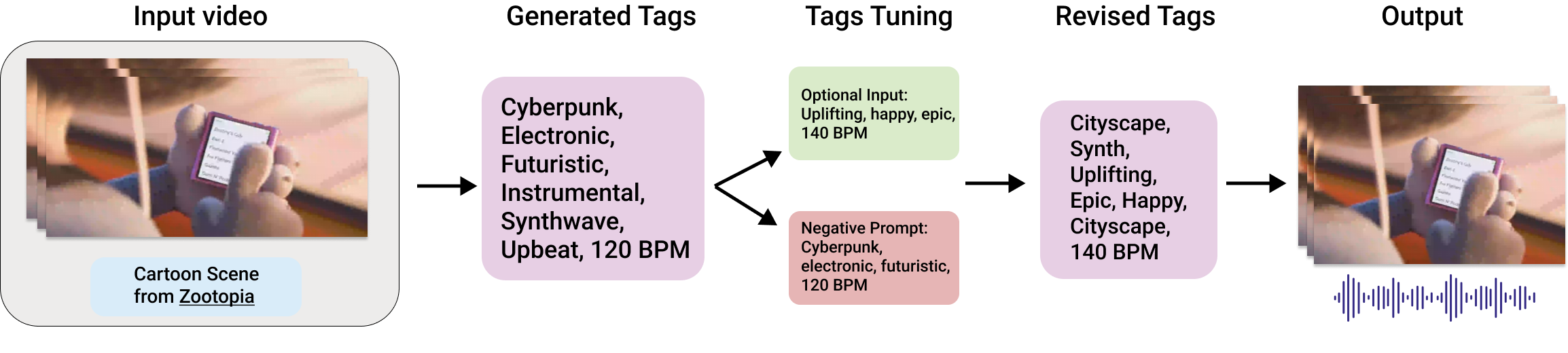}}
\caption{The process of generating music with SONIQUE. First, the input video (e.g., a cartoon scene from Zootopia) is analyzed to generate descriptive tags such as "Cyberpunk, Electronic, Futuristic, Synthwave, Upbeat, 120 BPM." Users can then fine-tune the music generation by providing additional prompts or specifying negative prompts. The final output is background music that matches both the video and user preferences.}
\label{fig:ft}
\end{figure*}

\noindent \textbf{Conditioned Audio Generation Model.} The foundation of SONIQUE's multi-model architecture is the conditional latent diffusion model from Stable Audio’s framework \cite{evans2024fast}. It constists of a VAE autoencoder \cite{kingma2022autoencoding}, a text encoder (CLAP) \cite{elizalde2022clap}, and a diffusion U-Net model. The VAE can compress 44.1kHz stereo audio into a latent space that enables faster generation and training time compared to working with raw audio samples \cite{evans2024fast}. This reduces computation requirements and increases efficiency. Differently than Stable Audio, we use a pre-trained CLAP \cite{elizalde2022clap} text encoder as it is difficult to have such large in-house text-audio dataset, and because LION-CLAP \cite{laionclap2023} (the pre-trained model we use), has shown great performance in multiple audio-text tasks \cite{soundsimilarity2024}. 
 The diffusion U-Net setup follows Stable Audio \cite{evans2024fast} to ensure the output meets professional quality and aligns with the input prompts. 

 Same as Stable Diffusion, the model we use here accepts \textit{negative prompts}, i.e. prompts that specify what the user does not want in the generation \cite{evans2024fast}. This feature ensures flexibility in our model and the capacity for the user to correct undesired or inaccurate tags generated from the video description. Figure \ref{fig:ft} shows the pipeline for a user to create music for a video.

\noindent \textbf{Training with Unpaired Data.} Dealing with paired audio-visual data not only limits the scalability of training but also reduces the diversity of the dataset, making it challenging for the model to learn to generate a wide range of music genres and instruments. In contrast, by leveraging descriptive tags, we can train with any music dataset that includes textual descriptions, providing far greater flexibility and variety. This was the approach we adopted with SONIQUE, where video data was used exclusively for evaluating the consistency of the generated music and for selecting the appropriate LLM, alongside other factors like model complexity. As a downside, timing alignment between video and music could be less precise, especially for complex or long sequences, making SONIQUE better suited for shorter video clips. 

\begin{figure*}[htb] 
\centerline{\includegraphics[width=18cm]{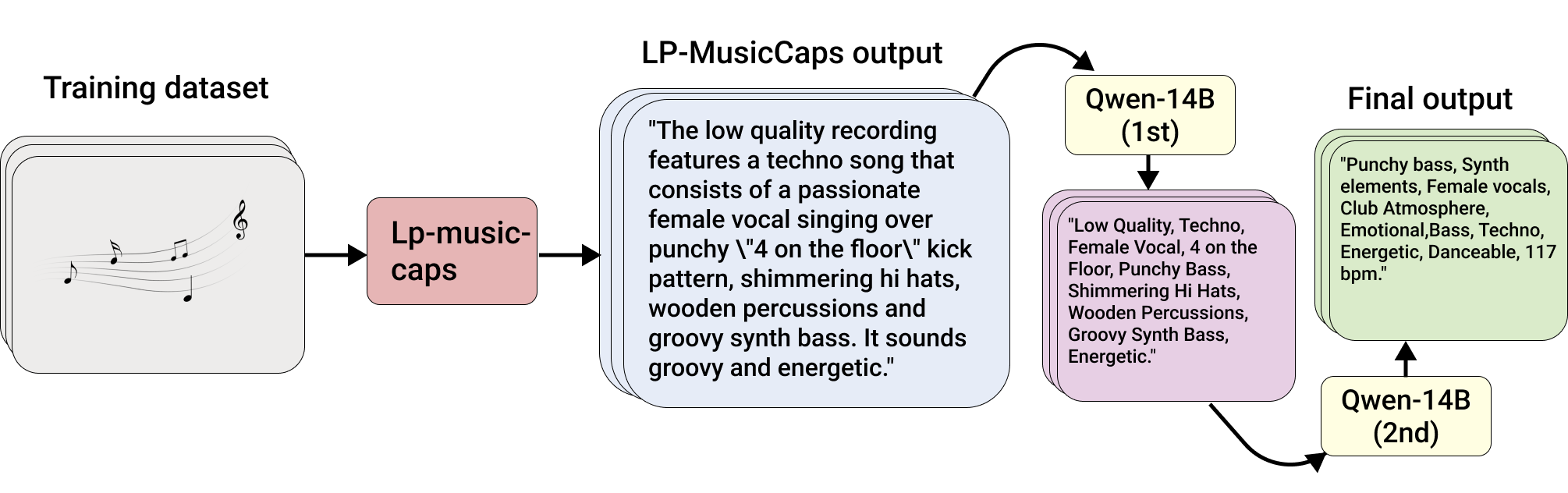}}
\caption{
In SONIQUE, tag generation for training starts by feeding raw musical data into LP-MusicCaps \cite{doh2023lpmusiccaps} to generate initial captions. These captions are processed by Qwen 14B \cite{bai2023qwen} in two steps: first, it converts the captions into tags, then it cleans the data by removing any incorrect or misleading tags (e.g., "Low Quality"). This results in a clean set of tags for training.}
\label{fig:llm}
\end{figure*}

As most of our music data does not have text descriptions, we tag them using lp-music-caps \cite{doh2023lpmusiccaps}, an available music understanding model. We tag music aspects such as instruments and genre. For estimating tempo we use librosa \cite{mcfee2015librosa}. The typical output from lp-music-caps 
is too verbose and rogue (using terms such as ``ethereal'' and ``soft''). 
To streamline the metadata, we employed Qwen 14B \cite{bai2023qwen} transform captions into concise, non-repetitive tags, as shown in Figure \ref{fig:llm}. To keep every metadata entry consistent, we randomly select up to eight tags from tracks with many tags, while also training on tracks with fewer than eight. Our setup ensures flexibility, and the balance between detail and model trainability. 


To connect the video to the music generation model, we convert the video description into the same type of tags, as described before.  
For that, we initially utilized Video-LLaMA \cite{zhang2023videollama} to perform caption to tags generation. The result tags were suboptimal in quality and precision when benchmarked against other LLMs. While 72B models offered robust performance, they were computationally expensive for our use case. We then explored several small models, including Gemma-7B \cite{google2024gemma}, LLaMA-7B \cite{touvron2023llama}, Mistral-7B \cite{inferless2024mistral} and Qwen 14B \cite{bai2023qwen}. Although LLaMA-7B was particularly vague, the other models showed similar outputs, suitable for our generation model. 

\noindent \textbf{Datasets and training.} Our dataset contains approximately 2644 hours of audio, which includes mix tracks and isolated stems from Slakh2100 \cite{manilow2019cutting}, Candombe \cite{Nunes2015}, GuitarSet \cite{xi2018guitarset}, GTZAN Genres \cite{tzanetakis_essl_cook_2001}, The MAESTRO Dataset \cite{hawthorne2018enabling}. Those available were chosen because of their quality and notable information about their genres or instruments. In addition, our dataset contains 745 hours of royalty-free music from Pixabay.

We first use the pretrained Descript-Audio-Codec autoencoder \cite{kumar2023highfidelity}, which was trained on a large dataset compiled of speech, music, and environmental sounds, but the output results in low quality. Inspired by Stable Audio's methodology \cite{evans2024fast}, we train the autoencoder on our dataset for 240,000 steps. 

The diffusion model was trained on 44.1kHz audio with 1,048,576 samples (23.8 seconds) due to computational and dataset limitations. The configuration follows Stable Audio's setup, it was trained for 140,000 steps with a batch size of 16 on four RTX8000 GPUs. Audio data was resampled to 44.1kHz and sliced to the required sample length during the training loop. Files longer than this length were cropped from a random starting point, while shorter files were padded with silence at the end.



\section{Evaluation}
\label{sec:majhead}


\noindent \textbf{Evaluation Metrics. }We employed three key quantitative metrics to evaluate the quality and relevance of the generated music: FDopenl3, KLpasst, and CLAP score, as outlined in Stable Audio's work \cite{evans2024fast}. In a nutshell, FDopenl3 measures the similarity between generated and reference audio sets, where lower values indicate closer resemblance; KLpasst evaluates the semantic correspondence between lengthy audio clips; and CLAP score assesses how well the generated audio aligns with a given text prompt. Our evaluation also utilizes code provided by Stable Audio \cite{evans2024fast} with the provided pre-computed statistics and embeddings with MusicCaps dataset \cite{agostinelli2023musiclm}.

\begin{figure}[htb]
\centering
\centerline{\includegraphics[width=9.5cm]{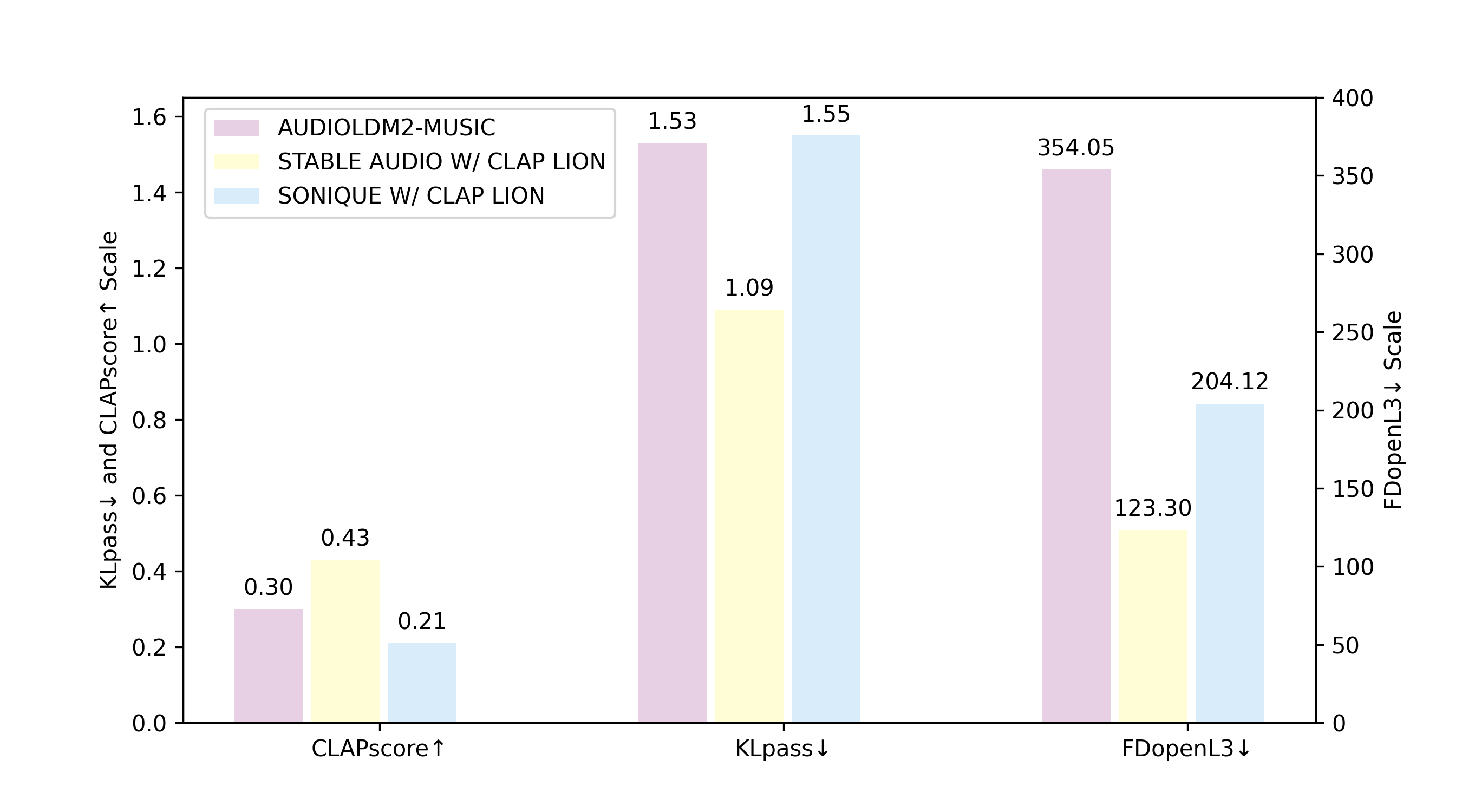}}
\caption{Quantitative results on MusicCaps. Other results are from  \cite{evans2024fast}.}
\label{fig:eval-metric}
\end{figure}

 We first generates 23 seconds audio with tags derived from the MusicCaps dataset \cite{agostinelli2023musiclm} and compare with the results from Stable Audio \cite{evans2024fast} and AudioLDM2-Music \cite{liu2023audioldm}. With only 2644 hours of musical data for training and mostly automatically generated tags, compared to 19.5k hours of good quality music and tags used in Stable Audio, 
 our model performed very competitively, particularly on FDopenl3, indicating it effectively captures the characteristics of clean, instrumental music (i.e. similar enough to MusicCaps). The KLpasst score suggests good semantic understanding, though there is room for refinement, while the CLAP score highlights potential improvements in aligning the music more closely with specific text prompts. This can be achieved by fine tuning the CLAP encoder, which was out of the scope of this work.

\begin{figure}[htb]
\begin{minipage}[b]{1.0\linewidth}
\centering
\centerline{\includegraphics[width=7cm]{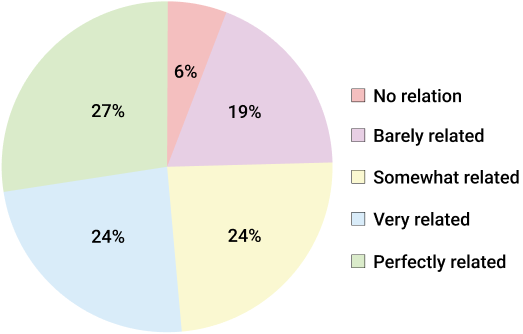}}
\end{minipage}
\caption{Human evaluation overall score}
\label{fig:eva}
\end{figure}

\noindent \textbf{Subjective evaluation.} We generate a demo with seven examples using SONIQUE. These generated videos were evaluated by a group of 38 individuals, including of artists with video editing backgrounds and music technology students. The evaluation followed a scale 1-5, rating how related was the music to the video:  No relation (1), Barely related (2), Somewhat related (3), Very related (4) and perfectly related (5). Figure \ref{fig:eva} shows overall ratings of though the seven demo videos. 


Overall, 75\% of users rated the generated audio as somewhat, very, or perfectly related to the video, with "perfectly related" being the most common rating. This positive feedback highlights SONIQUE's effectiveness in producing audio that aligns well with video content. However, 25\% of users found the audio to have little or no relation to the video, indicating that the model struggles to capture the mood or sync the music with specific video events. This suggests a need for improvement, particularly in the caption-to-tags stage, to enhance the model’s ability to generate more timely accurate audio.


\section{Conclusion}

SONIQUE offers a flexible, open-source, and scalable solution for video-to-music generation by removing the need for paired datasets. It uses LLMs to convert video descriptions into musical tags, combined with a U-Net-based diffusion model for customizable music generation. The model was trained on a diverse dataset of 2,644 hours of royalty-free music, tagged using LP-MusicCaps, ensuring a wide variety of genres and instruments. While the model performs well in generating clean, instrument-based music with good semantic understanding, it faces limitations in precise timing alignment, particularly for longer video clips, which we will address in future work. 

\bibliographystyle{unsrt}
\bibliography{references}

\end{document}